\documentclass[aps,showpacs,nofootinbib]{revtex4}
\usepackage{amsmath,amscd,amssymb}
\usepackage{color}
\usepackage{latexsym}
\usepackage{mathrsfs}
\usepackage{graphicx}
\usepackage{bbm}      

\newcommand{\dslash}{\partial \hskip -0.6em /}
\newcommand{\Dslash}{D \hskip -0.7em /}
\newcommand{\zr}[1]{\mbox{\hspace*{#1em}}}
\newcommand{\ID}{\mbox{{\sf 1}\zr{-0.16}\rule{0.04em}{1.55ex}\zr{0.1}}}
\newcommand{\vek}{\mathbf}

\begin{document}

\baselineskip18pt

\title{Quantum Stabilization of Cosmic Strings}

\author{H. Weigel}

\affiliation{Physics Department, Stellenbosch University,
Matieland 7602, South Africa}

\author{M. Quandt}
\affiliation{Institute for Theoretical Physics, University of T\"ubingen,
D--72076 T\"ubingen, Germany}

\author{N. Graham}
\affiliation{Department of Physics, Middlebury College
Middlebury, VT 05753, USA}

\begin{abstract}
In the standard model, stabilization of  a classically unstable cosmic 
string may occur through the quantum fluctuations of a heavy fermion doublet.
We review numerical results from a semiclassical expansion in a reduced 
version of the standard model. In this expansion the 
leading quantum corrections emerge at one loop level for many internal 
degrees of freedom. The resulting vacuum polarization energy and the binding 
energy of occupied fermion energy levels are of the same order, and must 
therefore be treated on equal footing. Populating these bound states 
lowers the total energy compared to the same number of free fermions.
Charged strings are already stabilized for a fermion mass only somewhat
larger than the top quark mass. Though obtained in a reduced version these results 
suggest that neither extraordinarily large fermion masses nor unrealistic couplings 
are required to bind a cosmic string in the standard model. Furthermore we also 
review results for a quantum stabilization mechanism that prevents closed 
Nielsen--Olesen type  strings from collapsing.

\end{abstract}

\pacs{11.27.+d, 03.65.Ge, 14.65.Jk}

\maketitle

\section{Introduction}

In this brief review article we summarize
some recent results on how quantum effects can stabilize 
classically unstable string--like configurations in the electroweak standard model. 
Such configurations are the particle physics analogs of vortices or magnetic flux 
tubes in condensed matter physics. They are usually called {\it cosmic strings} to 
distinguish them from the fundamental variables in string theory, and also to 
indicate that they typically stretch over large length scales.  For electroweak 
strings or $Z$--strings\cite{Vachaspati:1992fi,Achucarro:1999it,Nambu:1977ag},
the $Z$--component of the electroweak gauge boson
acquires the  structure of an Abelian Nielsen--Olesen type  
vortex \cite{Nielsen:1973cs}.

Such strings may have emerged copiously at various epochs in the early universe, 
at interfaces between regions of different vacuum expectation values of the Higgs 
field(s) before electroweak symmetry breaking was reached. If these strings 
are stable, they should have survived and we should be able to observe them today.  
While their direct gravitational effects are negligible, $Z$--strings can still be 
relevant for cosmology at a sub--dominant level\cite{Achucarro:2008fn,Copeland:2009ga}.
Their most interesting consequences originate, however, from their coupling to the 
standard model fields. $Z$--strings provide a source for primordial magnetic 
fields\cite{Nambu:1977ag} and they also offer a scenario for baryogenesis with a 
second order phase transition\cite{Brandenberger:1992ys,Brandenberger:1994bx}.
In contrast, a strong first order transition, as required by the usual bubble
nucleation scenario, is almost certainly ruled out in the electroweak standard 
model\cite{EWPhase} without non--standard additions such as supersymmetry or 
higher--dimensional operators\cite{Grojean:2004xa}. For reviews on cosmic 
strings and cosmological implications of networks of strings see {\it e.g.} 
Refs.~\cite{Vilenkin:1984ib,Hindmarsh:1994re,Brandenberger:1993by}.

There are numerical simulations showing that distributions or networks of
intersecting strings are stable and approach a so--called scaling limit, 
{\it cf.} Ref.~\cite{Albrecht:1989mk} and references therein. However, 
the standard model has neither a classical nor a topological mechanism  
to stabilize an isolated string to ensure that we would still
observe string networks at the present epoch. Thus these interesting
configurations  are only viable if they are bound dynamically by
their interaction  with the quantum  fluctuations of the fields. The
most important contributions  are expected to come from (heavy)
fermions, since their quantum energy dominates 
in the limit $N_C \to \infty$, where $N_C$ is the number of any other internal 
degree of freedom such as QCD color, for instance. The Dirac spectrum in typical 
string backgrounds is deformed to contain either an exact or near zero 
mode, so that fermions can substantially lower their energy by binding to 
the string. This binding effect can overcome the classical energy required 
to form the string background. However, the remaining spectrum of modes is 
also deformed and for consistency its contribution (the zero--point or
vacuum polarization energy) must be taken into account as well. Heavier fermions 
are expected to provide more binding since the energy gain per fermion charge is 
higher and the Yukawa coupling is stronger. A similar scenario is also suggested 
by decoupling arguments\cite{D'Hoker:1984ph}.   

Strings coupled to quantum fields have been investigated previously. 
Naculich\cite{Naculich:1995cb} has shown that fermion fluctuations destabilize 
the string in the limit of weak coupling. The quantum properties of $Z$--strings 
have also been connected to non--perturbative anomalies\cite{Klinkhamer:2003hz}.  
The emergence or absence of exact neutrino zero modes in a $Z$--string background 
and the possible consequences for the string topology were investigated in 
Ref.~\cite{Stojkovic}. A first attempt at a full calculation of the fermionic 
quantum corrections to the $Z$--string energy was carried out in 
Ref.~\cite{Groves:1999ks}. However, those authors were only able to compare the 
energies of two string configurations because of limitations arising from the 
non--trivial behavior at spatial infinity. We will present a solution to this 
serious obstacle.  The fermionic vacuum polarization energy of the simpler 
Abelian Nielsen--Olesen vortex has been estimated in Ref.~\cite{Bordag:2003at}
with a non--standard regularization that only subtracts the divergences within 
the heat--kernel expansion. Quantum energies of bosonic fluctuations in string 
backgrounds were also calculated in Ref.~\cite{Baacke:2008sq}. Finally, the 
dynamical fields coupled to the string can also result in (Abelian or non--Abelian) 
currents running along the core of the string. The time evolution of such 
structured strings was studied in Ref.~\cite{Lilley:2010av}, where the current 
was induced by the coupling to an extra scalar field.

This review is based on a number of 
publications\cite{Weigel:2009wi,Weigel:2010pf,Weigel:2010zk,Graham:2011fw} 
studying the contribution of fermion quantum corrections to the vacuum 
polarization energy of a straight and infinitely long cosmic 
string. The technical details of the underlying calculation are 
thoroughly described in the appendices of Ref.~\cite{Graham:2011fw}.

Cosmic strings normally carry magnetic flux and must either extend to spatial 
infinity or be closed. In the latter scenario, the straight strings described above 
would be replaced by quantum stabilized toroidal strings, or \emph{torons}. 
A simple application of Derrick's theorem \cite{Derrick:1964ww} reveals that torons 
are classically unstable, but this conclusion may again be reversed by quantum 
corrections. In section VI a simple application\cite{Quandt:2013qxa} of Heisenberg's 
uncertainty principle suggests that quantum fluctuations indeed prevent toroidal 
strings from collapsing.

This review is organized as follows: In section \ref{sec:model} we introduce the 
model that we use to investigate the interaction of a cosmic string with a fermion 
doublet. In section \ref{sec:scat} we explain the essential ingredients of the 
spectral methods that allow us to compute the vacuum polarization energy for 
string--like configurations that are translationally invariant with respect to 
one or more coordinates. We present numerical results for the vacuum energy of a 
string background in section \ref{sec:vacpol}, and discuss the binding of charged 
strings in sections \ref{sec:charged}. In section \ref{sec:var} we briefly 
report some novel results based on enlarging
the limited subset of parameters used in the previous numerical analysis. 
In section \ref{sec:closed} we review a quantum mechanical mechanism to stabilize
closed Nielsen--Olesen type strings finally we conclude in section \ref{sec:conclusion}.

\section{Model}
\label{sec:model}
For the current investigation the fermion doublet will be considered degenerate 
so that the introduction of a matrix notation for the Higgs field is appropriate. Two 
angles $\xi_1$ and $\xi_2$ that parameterize the isospin orientation\cite{Graham:2006qt} 
serve as variational parameters of the string configuration. The isospin group $SU(2)$ 
can, in general, be described by three Euler angles; in the considered configurations the 
third angle picks up the winding of the string in azimuthal direction. For notational 
simplicity, we introduce the abbreviations $s_i={\rm sin}(\xi_i)$ and $c_i={\rm cos}(\xi_i)$. 
Then the string configuration reads
\begin{equation}
\Phi=vf_H(\rho)
\begin{pmatrix}
s_1 s_2\, {\rm e}^{-in\varphi} & -ic_1-s_1c_2 \\[2mm]
-ic_1+s_1c_2 & s_1 s_2\, {\rm e}^{in\varphi}
\end{pmatrix}
\label{eq:Higgs}
\end{equation} 
for the Higgs field and 
\begin{equation}
\vec{\vek{W}} = \vek{W}^a\,\frac{\sigma^a}{2} = n\,s_1\,s_2\,
\frac{f_G(\rho)}{g\rho}\,\vek{e}_\varphi\,
\begin{pmatrix}
s_1 s_2 & -\left(ic_1+s_1c_2\right){\rm e}^{in\varphi} \\[2mm]
\left(ic_1-s_1c_2\right){\rm e}^{-in\varphi} & -s_1s_2
\end{pmatrix}
\label{eq:gauge}
\end{equation}
for the gauge boson (in temporal gauge).\footnote{The gauge field is a vector 
both in coordinate and isospin space, and the generators of 
the isospin algebra are $\sigma^a/2$ with the Pauli-matrices $\sigma^a$.} 
The variables $\rho$ and $\varphi$ are polar coordinates in the plane perpendicular 
to the string, while the Higgs vacuum expectation value~$v$ and the gauge coupling 
constant $g$ are model parameters. The string configuration involves profile 
functions $f_H$ and $f_G$ which are the analogs of the Nielsen--Olesen vortex 
profiles with boundary conditions
\begin{align}
f_G\,,\,f_H\, &\to\, 0 
\quad {\rm for} \quad \rho\,\to\, 0
\\[2mm]
f_G\,,\,f_H\, &\to\, 1
\quad {\rm for} \quad \rho\,\to\, \infty\,.
\label{eq:bc}
\end{align}
The integer $n$ is the winding number of the string, for which we 
will typically take $n=1$ in the numerical calculations. 

This simplified version of the standard model has a vanishing Weinberg angle so 
that the $U(1)$ hypercharge decouples and the gauge symmetry is $SU(2)_L$. Then 
the (classical) boson part of the Lagrangian reads
\begin{equation}
\mathcal{L}_{\rm bos}=-\frac{1}{2} {\rm tr}\left(G^{\mu\nu}G_{\mu\nu}\right) 
+\frac{1}{2} {\rm tr} \left(D^{\mu}\Phi \right)^{\dag} D_{\mu}\Phi
- \frac{\lambda}{2} {\rm tr} \left(\Phi^{\dag} \Phi - v^2 \right)^2 \,,
\label{eq:Lboson}
\end{equation}
with the covariant derivative $D_\mu = \partial_\mu - i g W_\mu$ and the 
$SU(2)$ field  strength tensor 
\begin{equation}
G_{\mu\nu}=\partial_\mu\,W_\nu-\partial_\nu W_\mu-ig\left[W_\mu,W_\nu\right]\,.
\label{fieldtensor}
\end{equation}
The classical boson masses are determined from $g$ and $v$ and the Higgs self--coupling 
$\lambda$ as $m_{\rm W}=gv/\sqrt{2}$ and $m_{\rm H}=2v\,\sqrt{\lambda}$ for the 
gauge and Higgs bosons, respectively\footnote{The three gauge bosons are degenerate
due to the vanishing Weinberg angle.}. The interaction with the fermion doublet is 
described by the Lagrangian
\begin{equation}
\mathcal{L}_{\rm fer}=i\overline{\Psi}
\left(P_L \Dslash  + P_R \dslash \right) \Psi
-f\,\overline{\Psi}\left(\Phi P_R+\Phi^\dagger P_L\right)\Psi
\label{eq:Lfermion}
\end{equation}
with the left/right--handed projector $P_{R,L}=\left(\ID\pm\gamma_5\right)/2$.
Upon spontaneous symmetry breaking the Yukawa coupling $f$ induces a fermion 
mass~$m=vf$. Assuming a heavy fermion doublet with
the mass of the top quark, the  standard model suggests the parameters
\begin{equation}
g=0.72\,,\qquad
v=177\,{\rm GeV}\,,\qquad
m_{\rm H}= 140\,{\rm GeV}\,,\qquad
f=0.99\,.
\label{eq:parameters}
\end{equation}
In the numerical search for a stable string, other model parameters are considered 
as well, \emph{cf.}~section \ref{sec:charged}.

\medskip\noindent
The classical energy per unit length of the string is determined
by $\mathcal{L}_{\rm bos}$,
\begin{equation}
\frac{E_{\rm cl}}{m^2}=2\pi\int_0^\infty \rho\, d\rho\,\Biggl\{
n^2s_1^2 s_2^2\,\biggl[\frac{2}{g^2}
\left(\frac{f_G^\prime}{\rho}\right)^2
+\frac{f_H^2}{f^2\rho^2}\,\left(1-f_G\right)^2\biggr]
+\frac{f_H^{\prime2}}{f^2}
+\frac{\mu_H^2}{4f^2}\left(1-f_H^2\right)^2\Biggr\}\,,
\label{eq:Ecl}
\end{equation}
where the dimensionless radial integration variable is related to the 
physical radius by $\rho_{\rm phys}=\rho/m$, and we have introduced
the mass ratio $\mu_{H}\equiv m_{\rm H}/m$. 

The central object of the present investigation is the fermion contribution 
to  the energy. It is obtained from the solutions to the Dirac equation in 
the two--dimensional plane perpendicular to the string (in a basis with 
$\gamma_0$ diagonal)
\begin{equation}
H\Psi_n=\omega_n \Psi_n 
\qquad {\rm with} \qquad
H=-i\begin{pmatrix}0 & \vec{\sigma}\cdot\hat{\rho} \cr
\vec{\sigma}\cdot\hat{\rho} & 0\end{pmatrix} \partial_\rho
-\frac{i}{\rho}\begin{pmatrix}0 & \vec{\sigma}\cdot\hat{\varphi} \cr
\vec{\sigma}\cdot\hat{\varphi} & 0\end{pmatrix} \partial_\varphi
+m\begin{pmatrix} 1 & 0 \cr 0 & -1\end{pmatrix} +H_{\rm int}\,,
\label{eq:Dirac}
\end{equation}
where the single particle Hamiltonian $H$ is extracted from 
$\mathcal{L}_{\rm fer}$, eq.~(\ref{eq:Lfermion}). Details of the 
interaction part, $H_{\rm int}$, will be discussed later, in particular 
concerning the subtleties at spatial infinity. The spectrum of the Dirac operator 
consists of bound state solutions with discrete eigenvalues $\omega_n=\epsilon_j$ and 
continuous scattering solutions whose eigenvalues $\omega$ are labeled by momentum $k$, 
i.e.~$\omega=\sqrt{k^2+m^2}$. Note also that the Dirac Hamiltonian 
for the string background anti--commutes with $\alpha_3=\gamma^0\gamma^3$ so that
the spectrum is charge conjugation invariant and it suffices to consider
the non--negative eigenvalues.

\section{Formalism}
\label{sec:scat}
The vacuum polarization energy is the renormalized sum of the changes of the 
zero--point energies of fermions\footnote{For bosons the overall sign needs
to be reversed.} in the background of a static configuration
\begin{equation}
E_{\rm vac}=-\frac{\hbar}{2}\sum_n\left(\omega_n-\omega_n^{(0)}\right)
\Bigg|_{\rm ren}
=-\frac{\hbar}{2}\sum_j \epsilon_j -
\hbar \int_0^\infty dk\, \omega_k\,\Delta\,\rho_{\rm ren}(k)\,.
\label{eq:def}
\end{equation}
Here $\omega_n$ are the energy eigenvalues in the presence of the string as obtained 
from the Dirac equation (\ref{eq:Dirac}), and $\omega_n^{(0)}$ are their free 
counterparts. In the second part of equation~(\ref{eq:def}) the changes of the 
single particle energies are split up into the contribution from distinct bound 
states ($\epsilon_j$) and an integral over the energies of the continuous 
scattering states weighted by the (renormalized) change of the density of scattering 
states, $\Delta\,\rho_{\rm ren}(k)$. This change is induced by the background fields 
and its determination will be outlined in the next subsection. The scattering states 
obey the standard dispersion relation $\omega_k=\sqrt{k^2+m^2}$. In eq.~(\ref{eq:def}), 
the factor $\hbar$ has been made explicit to stress that $E_{\rm vac}$ is a quantum 
effect. Unless stated otherwise, natural units ($\hbar=1$ and $c=1$) will be adopted 
in the remainder of this paper.

The computation of the fermion contribution to the vacuum polarization energy 
of a cosmic string proceeds in three stages. First, spectral methods are 
employed to express this energy in form of scattering data\cite{Graham:2009zz} 
while maintaining standard renormalization conditions. Second, this approach 
is extended to accommodate configurations that are translationally invariant 
in one or more spatial directions, using the so--called interface 
formalism\cite{Graham:2001dy}. Third, the non--trivial structure of the 
cosmic string configuration at spatial infinity must be addressed. More 
precisely, since the fields approach a pure gauge rather than zero at spatial 
infinity, the immediate application of spectral methods is impossible. Instead, a 
particular local gauge transformation is performed, which facilitates 
the formulation of a well--behaved scattering problem\cite{Weigel:2010pf}.

\subsection{Spectral Methods}

The basic idea of the spectral approach is to express 
$\Delta\,\rho_{\rm ren}(k)$ as the momentum derivative of the phase shifts 
of the scattering states\cite{Krein:1953}. Then
\begin{equation}
E_{\rm vac}=-\frac{1}{2}\sum_j \epsilon_j -
\sum_{\ell} D_\ell \int_0^\infty \frac{dk}{2\pi}\, \omega_k\, 
\frac{d}{dk}\left[\delta_\ell(k)\right]_{\rm ren}\,.
\label{eq:attempt1}
\end{equation}
In this formulation, it is implicitly assumed that the system has enough 
symmetry to allow for a partial wave decomposition with a degeneracy factor 
$D_\ell$ for the partial wave of (generalized) angular momentum~$\ell$. 
The major concern in eq.~(\ref{eq:attempt1}) is renormalization. As it stands, 
the vacuum polarization energy eq.~(\ref{eq:attempt1}) in three spatial 
dimensions is quadratically divergent at large momenta. In this regime, 
however, the Born series adequately represents the phase shift. It 
is therefore advantageous to introduce
\begin{equation}
E^{(N)}=-\frac{1}{2}\sum_j \epsilon_j -
\sum_{\ell} D_\ell \int_0^\infty \frac{dk}{2\pi}\, \omega_k\,
\frac{d}{dk}\left[\delta_\ell(k)\right]_N\,,
\label{eq:attempt2}
\end{equation}
where the subscript on the phase shift indicates the subtraction of 
the $N$ leading terms of the Born series to the phase shift, 
$\left[\delta_\ell\right]_N\equiv\delta_\ell-\delta_\ell^{(1)}-\delta_\ell^{(1)}
-\ldots-\delta_\ell^{(N)}$.
Choosing $N$ sufficiently large renders $E^{(N)}$ finite. 

Technically the Born series is an expansion in powers of $H_{\rm int}$. The contribution
of any such power to the vacuum polarization energy can be associated with
a Feynman diagram
\begin{equation}
\sum_{\ell} D_\ell
\int_0^\infty \frac{dk}{2\pi}\, \omega_k\, \frac{d}{dk}
\delta_\ell^{(n)}(k)\quad \mbox{\LARGE $\sim$}\qquad
\begin{minipage}{3.5cm}\vspace{-0.3cm}\includegraphics[width=15mm, height=15mm]{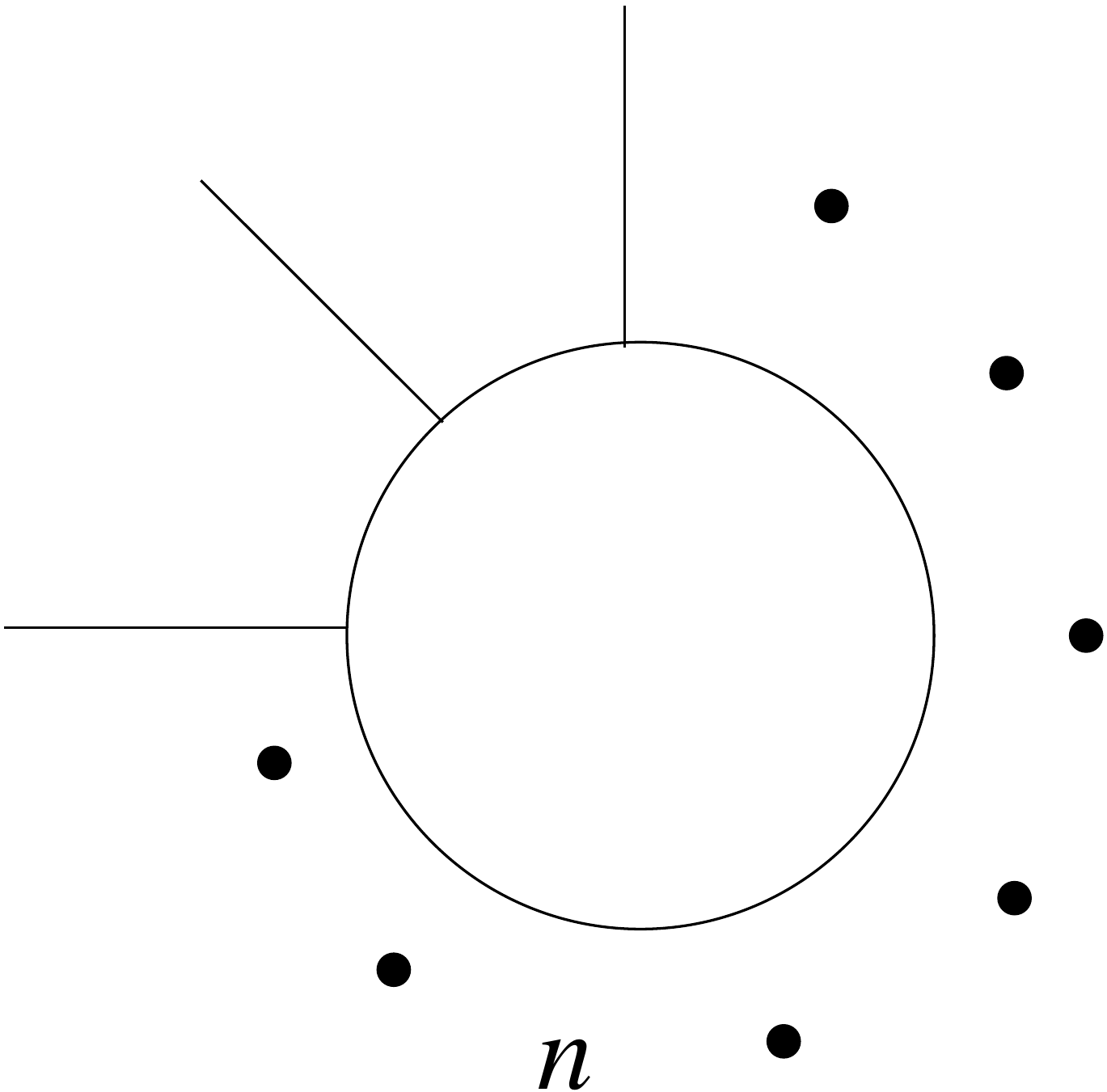}\end{minipage}
\,,
\label{eq:EbornFD}
\end{equation}
where the loop corresponds to the quantum fluctuation, and the external lines
represent insertions of the background field, {\it i.e.} the interaction of
the cosmic string with the fermions. Therefore, the Born subtraction in 
equation~(\ref{eq:attempt2}) corresponds to the sum $E^{(N)}_{\rm FD}$ of all 
Feynman diagrams with up to $N$ insertions of the background interaction. These 
diagrams can be computed with standard techniques; in particular dimensional 
regularization can be implemented to handle the ultra--violet divergences. 
It must be emphasized at this point that the use of the Born series does not 
imply any approximation for the phase shift; it is always computed exactly
(eventually numerically).  The Born series merely serves to identify the 
divergences at large momenta in the spectral approach.

The main advantage of transferring of the divergences into Feynman 
diagrams is that they can now be straightforwardly combined with the 
counterterm contribution to the energy, $E_{\rm CT}$. The latter is found by 
substituting the background configuration into the counterterm Lagrangian in 
just the same way as the classical energy, eq.~(\ref{eq:Ecl}) is obtained
from $\mathcal{L}_{bos}$. In any (multiplicatively) renormalizable theory, 
the counterterm Lagrangian has the same structure as the classical one and a 
suitable choice of its coupling constants cancels all ultra--violet divergences 
completely.  Then $E^{(N)}_{\rm FD}+E_{\rm CT}$ is free of any divergences. The finite 
pieces of the counterterm coupling constants will be uniquely determined from 
appropriate conditions describing properties of the particles that are 
associated with the fields. This procedure will be briefly discussed in 
section.~\ref{subsec:Onshell}.  In total, the sum
\begin{equation}
E_{\rm vac}=E^{(N)}+E^{(N)}_{\rm FD}+E_{\rm CT}
\label{eq:Efinite}
\end{equation}
gives an unambiguous result for the vacuum polarization energy once the 
renormalization conditions are fixed. A first principle derivation of this 
result, based on a quantum field theoretic formulation of the energy momentum 
tensor and the analytic properties of the Greens function, is presented in 
Ref.~\cite{Graham:2002xq}. We stress that eq.~(\ref{eq:Efinite}) is a sum of 
separately finite contributions, which does not involve any large number 
(\emph{e.g.} a cut--off).

\subsection{Interface Formalism}

Equation~(\ref{eq:Efinite}) is perfectly suited for a computation of the vacuum 
polarization energy for static configurations that allow for a full decomposition
into partial waves. However, the cosmic string does not exhibit a three--dimensional 
rotational invariance; rather it is translationally invariant along the $z$--axis.
In this scenario the wave--function of the quantum fluctuation factorizes into 
\begin{equation}
\Psi(\vec{x},t)\sim {\rm e}^{-i\omega t} \, {\rm e}^{ipz}\,
\psi_{k}(\vec{\rho}) .
\label{eq:Inface1}
\end{equation}
Here $\psi_{k}(\vec{\rho})$ is the reduced wave--function for a two--dimensional 
scattering problem in the plane perpendicular to the string  with the total
dispersion relation $\omega=\sqrt{p^2+k^2+m^2}=\sqrt{p^2+\omega_k^2}$.
With the replacements $\epsilon_j\to\sqrt{p^2+\epsilon_j^2}$ and 
$\omega_k\to\sqrt{p^2+\omega_k^2}$ in equation~(\ref{eq:attempt2}), an 
integration over $p$ (with measure $dp/2\pi$) yields the vacuum polarization
energy \emph{per unit length} of the string. However, this procedure runs into 
an immediate problem: The scattering data do not depend on the momentum $p$ 
and hence the $p$--integral will not be finite for any $N$. A careful analysis 
must treat the $p$--integral in dimensional regularization\cite{Graham:2001dy}
\begin{equation}
E^{(N)}\sim \frac{\Gamma(-\frac{1+d}{2})}{2(4\pi)^{\frac{d+1}{2}}}
\sum_\ell D_\ell
\Bigg\{\sum_j\left(\epsilon_j\right)^{\frac{d+1}{2}} 
+ \int_0^\infty \frac{dk}{\pi} \left(k^2+m^2\right)^{\frac{d+1}{2}}\,
\frac{d}{dk}\left[\delta_\ell(k)\right]_{N}\Big]\Bigg\}\,,
\label{eq:Inface2}
\end{equation}
where $d$ is the analytic dimension of the subspace in which the configuration
is translationally invariant. The divergence now manifests itself via the 
singularity of the $\Gamma$--function coefficient as $d\to1$. Due to sum rules 
for scattering data\cite{Puff:1975zz}, which represent generalizations of 
Levinson's theorem \cite{Levinson:1949}, the expression in curly 
brackets in eq.~(\ref{eq:Inface2})
vanishes as $d\to1$ which cancels the $\Gamma$-function pole. Hence the limit 
$d\to1$ can indeed be taken\cite{Graham:2001dy}. The result is
\begin{equation}
E^{(N)}=\frac{1}{4\pi}\sum_\ell D_\ell \Bigg\{
\int_0^\infty \frac{dk}{\pi}
\left[\omega_k^2\, {\rm ln}\left(\frac{\omega_k^2}{\mu_r^2}\right)-k^2\right]\,
\frac{d}{dk}\left[\delta_\ell(k)\right]_N 
+\sum_j\left[\epsilon_j^2\, {\rm ln}\frac{\epsilon_j^2}{\mu_r^2}
-\epsilon_j^2+m^2\right]\Bigg\}\,,
\label{eq:Ereg}
\end{equation}
where $E^{(N)}$ is expressed through the non--negative part of the spectrum
thanks to charge conjugation invariance in the string background. Here
$\mu_r$ is an arbitrary renormalization scale that has no effect on $E^{(N)}$ by 
exactly the same sum rules. The expression (\ref{eq:Ereg}) for $E^{(N)}$ replaces 
the analog in equation~(\ref{eq:Efinite}). When comparing the two expressions,
it is apparent that the function multiplying the (Born subtracted) phase shift 
is of higher power in $k$ than its counterpart before integrating over the 
momentum $p$. Hence $N$ must be increased when further dimensions are added 
in which the configuration is translationally invariant. This, of course, 
merely reflects the fact that ultra--violet divergences become more severe in 
higher dimensions.

The expressions obtained so far for the vacuum polarization
energy rely strongly on the analytic properties of the scattering data. These properties 
also allow for numerical calculations using purely imaginary momenta. 
This has (at least) two advantages: (i) the oscillating 
phase shifts turn into exponentially decaying (logarithms of) Jost functions, 
and (ii) the momentum integral and the sum over angular momenta may be exchanged. 
While (i) drastically improves numerical stability, (ii) significantly simplifies 
the treatment of the logarithmic divergences that emerge at the third and fourth 
order of the Born and Feynman expansions. These technical details are discussed at 
length in Ref.~\cite{Graham:2011fw}; we address them briefly in section~\ref{sec_fake}.

\subsection{Choice of Gauge}

The results presented thus far establish a formalism for computing the renormalized 
vacuum polarization energy of background fields in the string geometry. 
However, there is stillthe problem that the string does not induce a 
well--behaved Born series when expanding in powers of $H_{\rm int}$ from 
eq.~(\ref{eq:Dirac}). Even though
the full Hamiltonian is gauge invariant, $H_{\rm int}$ is not and does not
vanish at spatial infinity. This obstacle appears because 
the Dirac Hamiltonian that is obtained by straightforward substitution
of the string background, eqs.~(\ref{eq:Higgs}) and~(\ref{eq:gauge}),
does not turn into the free Dirac Hamiltonian at $\rho\to\infty$. 
Instead it becomes
$H\to U^\dagger(\varphi) H_{\rm free} U(\varphi)$. This local gauge transformation 
is a consequence of the string winding and acts only on the left--handed fermions
\begin{equation}
U(\varphi)=P_L {\rm exp}\left(i\hat{n}(\varphi)\cdot\vec{\tau}\,\xi_1\right)+P_R
\qquad {\rm with} \qquad
\hat{n}(\varphi)=
\begin{pmatrix}
s_2\, {\rm cos}(n\varphi) \cr 
-s_2\, {\rm sin}(n\varphi) \cr c_2
\end{pmatrix}\,.
\label{eq:GT0}
\end{equation}
Unfortunately, the obvious gauge transformation 
$H \to U(\varphi)HU^\dagger(\varphi)$  does not solve the problem completely: 
Although it would generate vanishing interactions at infinity, it will also induce a 
$1/\rho^2$ potential at the core of the string, $\rho \to 0$. 
This might still yield well--defined phase shifts, though they would likely 
be difficult to compute numerically. In any event, the conditions underlying 
the analyticity of the scattering data are certainly violated by this
singular behavior. 
As argued at the end of the previous section, analyticity is central for numerical 
feasibility of our approach. As a solution, we can define a radially 
extended gauge transformation
\begin{equation}
U(\rho,\varphi)=P_L {\rm exp}\left(i\hat{n}\cdot\vec{\tau}\,\xi(\rho)\right)+P_R\,.
\label{eq:radialGT}
\end{equation}
This transformation fixes the gauge and leads to the interaction term 
in equation~(\ref{eq:Dirac})
\label{subsec:Gauge}
\begin{eqnarray}
H_{\rm int}&=&
m\left[\left(f_H{\rm cos}(\Delta)-1\right)
\begin{pmatrix} 1 & 0 \cr 0 &-1\end{pmatrix}
+if_H\,{\rm sin}(\Delta)\begin{pmatrix}0 & 1 \cr -1 & 0\end{pmatrix}
\hat{n}\cdot\vec{\tau}\right]
+\frac{1}{2}\frac{\partial \xi}{\partial \rho}
\begin{pmatrix}-\vec{\sigma}\cdot\hat{\rho}
& \vec{\sigma}\cdot\hat{\rho} \cr
\vec{\sigma}\cdot\hat{\rho}
& -\vec{\sigma}\cdot\hat{\rho}\end{pmatrix}\hat{n}\cdot\vec{\tau}
\nonumber \\[3mm]
&&\quad
+\frac{ns_2}{2\rho}\, \begin{pmatrix}-\vec{\sigma}\cdot\hat{\varphi}
& \vec{\sigma}\cdot\hat{\varphi} \cr
\vec{\sigma}\cdot\hat{\varphi}
& -\vec{\sigma}\cdot\hat{\varphi}\end{pmatrix}
\Big[f_G\,{\rm sin}(\Delta)\,I_G(\Delta)
+(f_G-1)\,{\rm sin}(\xi(\rho))\,I_G(-\xi)\Big]\,.
\label{eq:DiracInt}
\end{eqnarray}
The new gauge function $\xi(\rho)$ appears via the difference 
$\Delta(\rho)\equiv\xi_1-\xi(\rho)$ and the isospin matrices are 
\begin{equation}
\hat{n}\cdot\vec{\tau}=\begin{pmatrix}
c_2 & s_2 {\rm e}^{in\varphi} \\[2mm] 
s_2 {\rm e}^{-in\varphi} & -c_2 \end{pmatrix}
\qquad {\rm and} \qquad
I_G(x)=\begin{pmatrix}
-s_2{\rm sin}(x) & [c_2{\rm sin}(x)-i{\rm cos}(x)]\,{\rm e}^{in\varphi} \\[2mm]
[c_2{\rm sin}(x)+i{\rm cos}(x)]\,{\rm e}^{-in\varphi} &
s_2{\rm sin}(x)\end{pmatrix}\,.
\label{eq:IG}
\end{equation}
Imposing the boundary conditions $\xi(0)=0$ and $\xi(\infty)=\xi_1$ for the
new gauge function $\xi(\rho)$ defines a well--behaved scattering problem. The 
specific form of $\xi(\rho)$ is irrelevant (apart from its boundary conditions)
and must not have any influence on the quantum energy, since it merely 
parameterizes a gauge transformation. 

Note that the gauge transformation is single--valued at spatial infinity, 
$U(\infty,\varphi)=U(\infty,\varphi+2\pi)$. In this respect it differs from
the analogous problem of fractional fluxes in QED. In that case a similar choice 
of gauge would not be a remedy; rather the calculation 
of the vacuum polarization energy requires the introduction of a \emph{return flux} 
to arrive at a well--behaved scattering problem\cite{Graham:2004jb}.  
The return--flux approach can also be used for the present calculation, but it is 
much more laborious numerically\cite{Weigel:2009wi,Weigel:2010pf}.

\subsection{Fake Boson Field}
\label{sec_fake}

The idea of utilizing a fake boson field to simplify the treatment of
higher order divergences was first implemented in Ref.~\cite{Farhi:2001kh}.
The third and fourth order contribution from the Born series produce 
logarithmic divergences but the corresponding Feynman diagrams are 
very cumbersome to evaluate. On the other hand, these logarithmic
divergences are similar to the ones found in 
the second order vacuum polarization energy of a boson field fluctuating about 
a scalar potential. Matching its strength appropriately and recalling that
for imaginary momenta the momentum integral and the angular momentum sum 
may safely be exchanged allows to replace
\begin{equation}
\sum_\ell D_\ell \left[\frac{d}{dt}\nu_\ell(t)\right]_N\,\longrightarrow\,
\frac{d}{dt}\left[\sum_\ell D_\ell\left(
\nu_\ell(t)-\nu^{(1)}_\ell(t)-\nu^{(2)}_\ell(t)\right)
-\sum_{\ell}\bar{D}_{\ell}\bar{\nu}^{(2)}_\ell(t)\right]
\label{eq:FakeBos}
\end{equation}
under the integral in eq.~(\ref{eq:Ereg}). The quantity $E^{(N)}$ with this 
replacement will be called~$E_\delta$. (Overlined quantities refer to the 
bosonic scattering data).  The replacement eq.~(\ref{eq:FakeBos}) must, of course, 
be accompanied by the boson Feynman diagram~$E_{\rm B}$ so that the total 
vacuum polarization energy becomes
\begin{equation}
E_{\rm vac}=E_\delta+E_{\rm FD}^{\rm ren.}+E_{\rm B}^{\rm ren.}\,,
\label{eq:Master}
\end{equation}
where the superscript indicates the inclusion of the counterterm 
contributions. Each of the three terms on the right hand side of 
equation~(\ref{eq:Master}) is separately ultra--violet finite.
The advantage of eq.~(\ref{eq:Master}) and the replacement 
eq.~(\ref{eq:FakeBos}) is now obvious: Instead of fermionic 
contributions up to order $N=4$, we only need to compute second 
order fermionic and bosonic Feynman diagrams and terms in the 
corresponding Born series.

\section{Numerical Results for the Vacuum Polarization Energy}
\label{sec:vacpol}

Before discussing  numerical results it is worthwhile 
to return to the issue of the particular gauge transformation, 
eq.~(\ref{eq:radialGT}). The resulting invariance provides an excellent 
reliability test for the numerical procedure: Modifying the shape of $\xi(\rho)$ 
while keeping its boundary conditions fixed alters $E_\delta$ and 
$E_{\rm FD}^{\rm ren.}$ in eq.~(\ref{eq:Master}). However, these individual 
changes are bound to compensate each other\cite{Weigel:2010pf}. 
We discuss this test on the numerical results in section B below.

For $\xi_2\ne\frac{\pi}{2}$ the Hamiltonian acquires imaginary
parts, though it still remains hermitian. This complicates the computation
of the phase shifts because in--coming and out--going spherical waves
obey different differential equations. Therefore most of the
numerical results are limited to
the case $\xi_2=\frac{\pi}{2}$ as in ref.\cite{Graham:2011fw}. We will
present some novel results without that restriction at the end of this 
section.

As a convention, all numerical results and dimensionful parameters 
from here on will be measured in appropriate units of the (perturbative) 
fermion mass $m$.

\subsection{Variational Ans\"atze}

Despite the simplification in eq.~(\ref{eq:Master}), the numerical 
computation is still expensive. The scattering data are extracted from a 
multi--channel problem and for the final result to be reliable several hundred
partial wave channels must be included.  This numerical effort restricts 
the number of variational parameters that can be used to characterize the 
profile functions. In addition to $\xi_1$, we introduce three scale parameters 
$w_H$, $w_W$ and $w_\xi$ via the ans\"atze
\begin{equation}
f_H(\rho)=1-{\rm e}^{-\frac{\rho}{w_H}}\,,\qquad
f_G(\rho)=1-{\rm e}^{-\left(\frac{\rho}{w_G}\right)^2}\,,\qquad
\xi(\rho)=\xi_1\left[1-
{\rm e}^{-\left(\frac{\rho}{w_\xi}\right)^2}\right]\,.
\label{eq:Ansaetze}
\end{equation}
The scale $w_\xi$ parameterizes the shape of the gauge profile which should 
not be observable \emph{c.f.} the remark at the beginning of this chapter. 
The other properties of the profiles are chosen to keep $E_{\rm cl}$ regular.

We have also considered an exponential parameterization for the gauge
field 
\begin{equation}
f_G(\rho)=1-\left(1+\frac{\rho}{w_G}\right)
{\rm exp}\left(-\frac{\rho}{w_G}\right)\,,
\label{eq:ExpAnsatz}
\end{equation}
which yields a slightly better agreement with the original Nielsen--Olesen
profiles which minimize $E_{\rm cl}$ for $\xi_1=\pi/2$. No significant
difference in $E_{\rm vac}$ was found between these ans\"atze.

\subsection{Gauge Invariance}

We check gauge invariance by varying the width $w_\xi$ of the gauge profile, 
$\xi(\rho)$. A typical result is shown in table~\ref{tab:Res1}.
\renewcommand{\arraystretch}{1.4} 
\begin{table}[t]
\centering
\begin{tabular}{l|lll|l}
\toprule
$w_\xi$ &~ $E_{\rm FD}^{\rm ren.}$ & $E_\delta$ &
$E_{\rm B}^{\rm ren.}$ & $E_{\rm vac}$ \\
\colrule
0.5 &~ -0.2515\,\,\,\, & 0.3489\,\,\,\, & 0.0046\,\,\,\, & ~0.1020 \\
1.0 &~ -0.0655 & 0.1606 & 0.0032 & ~0.0983 \\
2.0 &~ -0.0358 & 0.1294 & 0.0038 & ~0.0974 \\
3.0 &~ -0.0320 & 0.1235 & 0.0056 & ~0.0971 \\
4.0 &~ -0.0302 & 0.1193 & 0.0080 & ~0.0971 \\
\botrule
\end{tabular}
\caption{\label{tab:Res1}
Numerical results for the various contributions~(\ref{eq:Master}) to
the fermion vacuum polarization energy in the minimal subtraction scheme.}
\end{table}
As expected, the individual contributions to $E_{\rm vac}$ depend strongly on 
$w_\xi$. However, these changes essentially compensate each other. Numerically 
the most cumbersome part of the calculation is $E_\delta$. From various 
numerical tests (changing the extrapolation scheme for partial wave sum, 
the momentum integration grid, etc.) its numerical accuracy is estimated to 
be at the 1\% level. Within that precision range, $E_{\rm vac}$ is independent 
of $w_\xi$, which verifies both gauge invariance and the equivalence of 
terms in Born and Feynman series. This precise equivalence of separately divergent 
quantities is fundamental to the spectral approach.

\subsection{On--Shell Renormalization}
\label{subsec:Onshell}

The results reviewed in the previous subsection were obtained in the $\overline{\rm MS}$ 
renormalization scheme, which essentially omits the non--divergent parts of the 
Feynman diagrams. In this scheme, the dependence of $E_{\rm vac}$ on the model 
parameters factorizes which simplifies the computation because this dependence can 
easily be traced from $E_{\rm cl}$. As a consequence, many parameter settings can be 
studied by rescaling instead of recomputing $E_{\rm vac}$. Any other scheme merely 
differs by manifestly gauge invariant (finite) counterterms. Since physically meaningful 
results require renormalization conditions that correspond to a particle interpretation, 
an additional (mild) parameter dependence in $E_{\rm FD}$ is induced. 

\begin{figure}[pb]
\centerline{
\includegraphics[width=7.0cm, height=4.5cm]{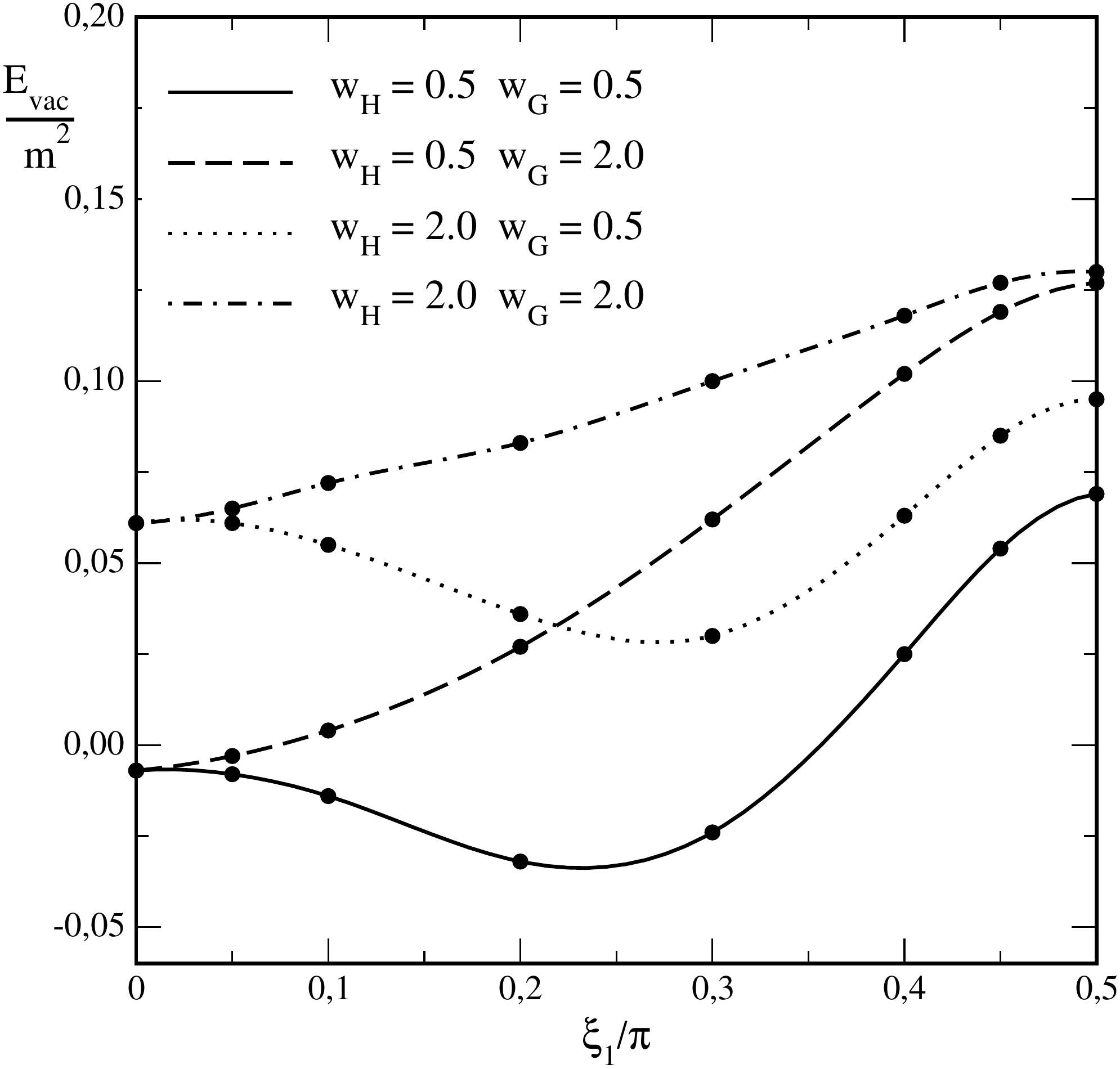}
\hspace{1.5cm}
\includegraphics[width=7.0cm, height=4.5cm]{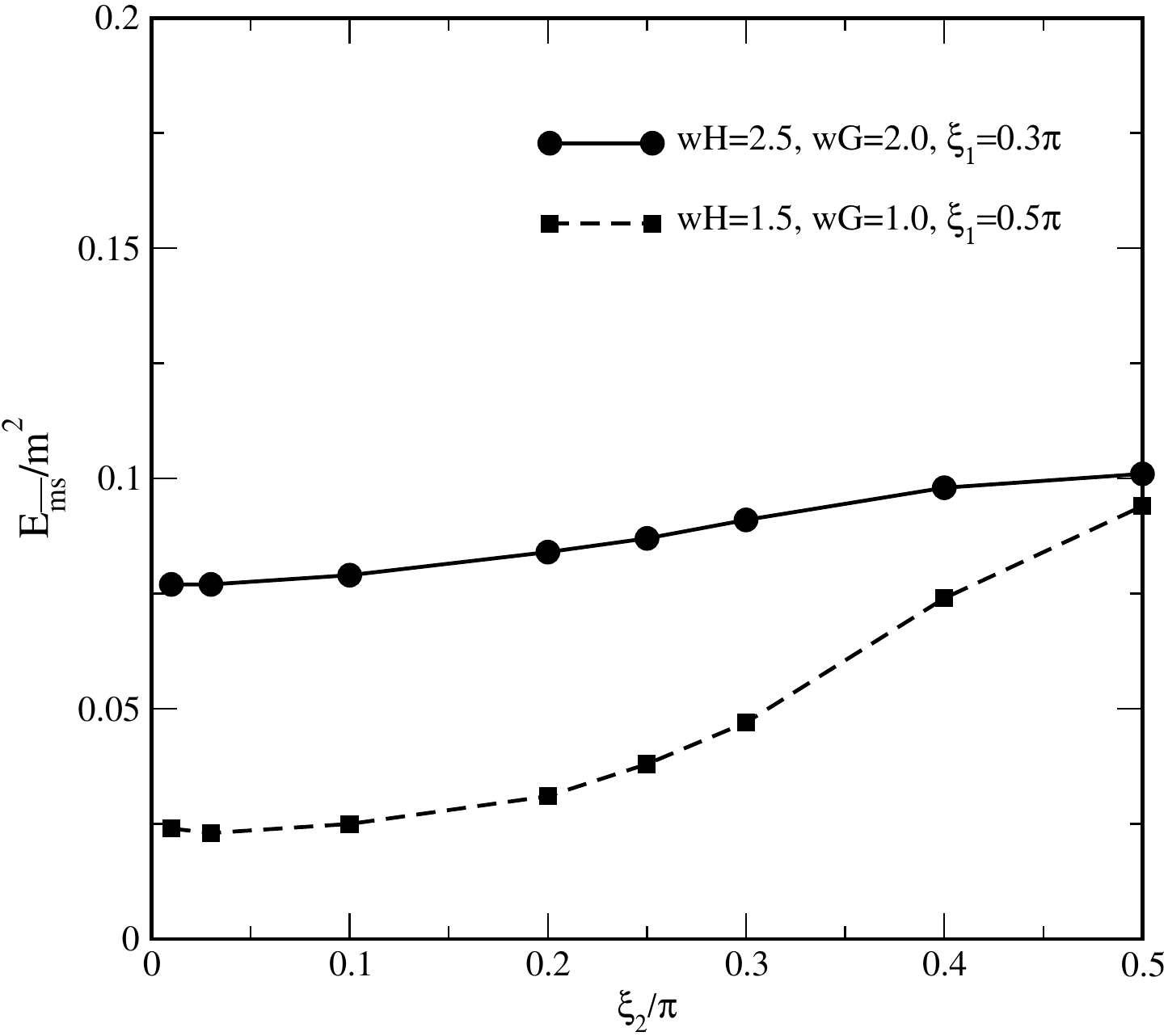}
}
\caption{\label{fig_onshell}Fermion vacuum polarization energy. Left panel:
on--shell renormalization scheme for configurations with $\xi_2=\frac{\pi}{2}$.
Right panel: some first results for arbitrary $\xi_2$ in the $\overline{\rm MS}$ scheme.}
\end{figure}

To be specific, we consider the so--called {\it on--shell} scheme, in which the 
coefficients of the four allowed counterterms (three terms from 
eq.~(\ref{eq:Lboson}) supplemented by ${\rm tr}\left[\Phi^\dagger\Phi\right]$) 
are determined such that
\begin{itemize}
\item[$\bullet$]
the tadpole graph vanishes implying that the Higgs \emph{vev} 
$v$ remains unchanged,
\item[$\bullet$]
the Higgs mass remains unchanged
\item[$\bullet$]
the normalization of one--particle states of the Higgs boson remains unchanged
\item[$\bullet$]
the normalization of one--particle states of the vector meson remains unchanged
\end{itemize}
in the presence of fermionic quantum corrections. Note that the vector meson 
mass~$M_W$ is not fixed by these conditions and thus will be a \emph{prediction} 
that includes quantum corrections. This suggests to tune the gauge coupling to 
reproduce an appropriate physical value for~$M_W\mbox{$\overset{!}{=}M_Z\approx90{\rm GeV}$}$. 
Typical results for the vacuum polarization energy per unit length of the string are 
shown in figure \ref{fig_onshell}, as functions of the variational parameters.
Except for narrow string configurations dominated by the Higgs field, the vacuum 
polarization energy turns out to be positive. Therefore, fermionic vacuum fluctuations 
alone do not provide any substantial binding and no stable uncharged string is found 
for the physically motivated parameters, eq.~(\ref{eq:parameters}), for which 
$E_{\rm cl}$ dominates the total energy. Since $E_{\rm cl}$ decreases quickly with 
increasing Yukawa coupling $f$, some stability is indeed seen for large $f$ and
narrow strings but in this regime the restriction to one fermion loop in the vacuum  
polarization energy is unreliable because of the unphysical Landau 
ghost contribution \cite{Ripka:1987ne,Hartmann:1994ai}.

\section{Charged Strings}
\label{sec:charged}

Cosmic strings induce many fermionic bound state levels for the two--dimensional 
scattering problem, whose energies are denoted by $\epsilon_j$. For 
$\xi_1=\xi_2=\pi/2$ there even exists an exact zero mode\cite{Naculich:1995cb}. In 
the three--dimensional setting these bound states acquire a longitudinal momentum 
for the motion along the symmetry axis, \emph{i.e.} their energies become
\begin{equation}
E_i(p_n)=\sqrt{\epsilon_i^2+p_n^2}
\qquad {\rm with} \qquad
p_n=\frac{n\pi}{L}\,.
\label{eq:E3d}
\end{equation}
Here $L$ is the length of the string. To leading order of the limit $L\to\infty$,
the sum over the discrete longitudinal momentum turns into a continuum integral, 
$\sum_n \,\longrightarrow\,\frac{L}{\pi}\int_{0}^{\infty} dp$. 
To find the minimal bound state 
contribution at a fixed charge (per unit length), a chemical potential $\mu$ with 
${\rm max}(|\epsilon_j|)\le\mu\le m$ is introduced, and all levels with $E_i(p)\le \mu$ 
are populated. This procedure defines a Fermi momentum for each level, 
$p^F_i(\mu)\equiv\sqrt{\mu^2-\epsilon_i^2}$ which enters the total charge per unit length 
of the string 
\begin{equation}
Q(\mu)=\sum_i \frac{p^F_i(\mu)}{\pi}\,.
\label{eq:TotalQ}
\end{equation}
This relation can be inverted to give $\mu=\mu(Q)$ and thus $p^F_i=p^F_i(Q)$. This 
quantity provides, for a given charge $Q$, the maximal momentum along the 
string direction for each fermion to get trapped in a two-dimensional bound state 
$\epsilon_i$ perpendicular to the string. 
From this the binding energy (per unit length) for a prescribed charge
\begin{equation}
E_{\rm bind}(Q)=\frac{1}{\pi}\sum_i\int_0^{p^F_i(Q)}dp \,
\left[\sqrt{\epsilon_i^2+p^2}-m\right]
\label{eq:Ebind}
\end{equation}
is computed relative to an equal number of free fermions that have 
energy~$m$ each.  

\begin{figure}[ht]
\centerline{
\includegraphics[width=7.5cm,height=6.0cm]{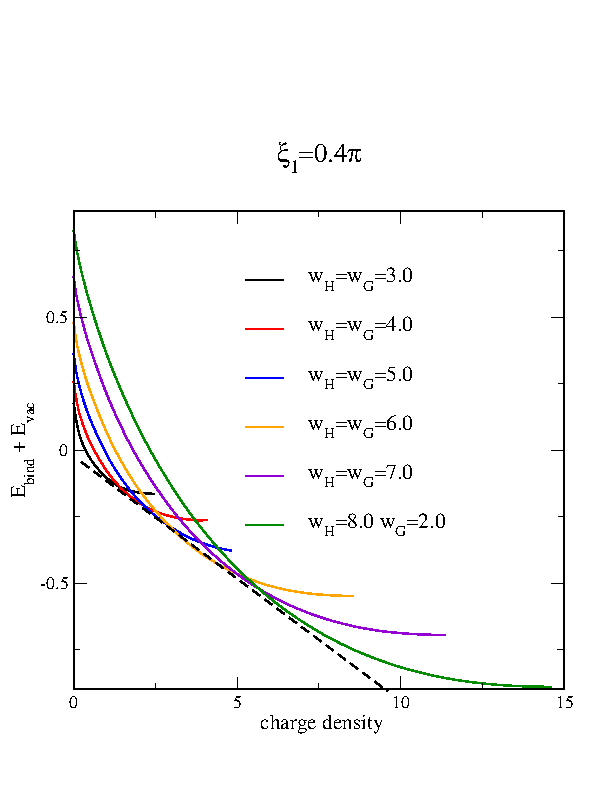}
\hspace{1cm}
\includegraphics[width=8.0cm,height=5.2cm]{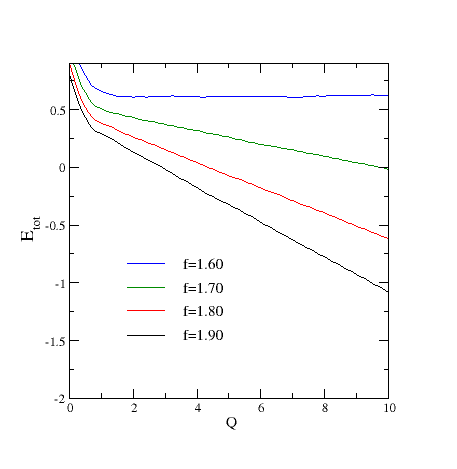}
}
\caption{\label{fig_ebline}
(Color online) Fermion contribution to the energy of a 
charged string (left panel). Total energy of the charged 
string (right panel) as function of the charge density.}
\end{figure}

Figure~\ref{fig_ebline} shows the fermion contribution to the binding energy, 
$E_{\rm vac}+E_{\rm bind}(Q)$. In the left panel the graph terminates for a 
given configuration when all available bound state levels are occupied 
and the charge cannot be increased any further. For small charges, narrow 
strings are favorable while the binding energy of wider strings decreases more 
quickly as $Q$ increases.
Surprisingly, the envelope along which $E_{\rm vac}+E_{\rm bind}$ is minimal 
(when varying the string parameters at fixed charge) forms a straight line. 
Extrapolating this line to~~$Q=0$~~indicates that the fermion vacuum polarization 
energy should (approximately) vanish. This extrapolation circumvents the Landau ghost 
problem mentioned in the previous section. 

To finally decide on dynamical stability, the classical energy must be included 
as well.  To this end several hundred configurations, that are characterized by 
the variational ans\"atze, eq.~(\ref{eq:Ansaetze}), were scanned. It is appropriate 
to label them by $s=1,2,\dots$ and to compute their total binding energy as
\begin{equation}
E^{(s)}_{\rm tot}(Q)=E^{(s)}_{\rm cl}
+N_C\left[E^{(s)}_{\rm vac}+E^{(s)}_{\rm bind}(Q)\right]
\label{eq:Ebindform}
\end{equation}
for a given charge. If
\begin{equation}
E_{\rm tot}(Q)\equiv {\rm min}_s\left[E^{(s)}_{\rm tot}(Q)\right]<0
\label{eq:Stable}
\end{equation}
a stable configuration has been found. The right panel of figure ~\ref{fig_ebline} 
shows $E_{\rm tot}$ as a function of charge for various values of the 
Yukawa coupling constant. Since the mass $m$ of a non--interacting fermion
sets the scale in the numerical analysis, this refers to different values 
of the dimensionless ratio $v/m$.
For $f\approx 1.6$ the classical and fermion energies essentially cancel 
each other and leave $E_{\rm tot}$ roughly independent of charge\footnote{The
strong dependence at small $Q$ is artificial because very narrow strings
have not been considered to avoid the Landau ghost inconsistency.}. Bound
objects are observed by further increasing the Yukawa coupling to about 
$f \approx 1.7$, which corresponds to a heavy fermion mass which is still less 
than twice the top quark mass. The minimizing configurations 
have $\xi_1\approx0$, {\it i.e.} they are dominated by the Higgs field.
The numerical results reviewed here largely support the existence of a novel
solution within the standard model of particle physics. These string 
configurations exist provided that a sufficient density of very heavy 
fermions is present which can be trapped to the string core to stabilize the 
configuration. Such conditions may have existed in the early universe, and the 
emerging cosmic strings should still be traceable today.  Inverting
this argument, not observing cosmic strings indicates that fermions 
coupling to the weak interactions in the usual way cannot have existed
with sufficient density if their mass exceeds about twice the
top--quark mass, unless they can decay to lighter fermions.

\section{Variational parameter $\mathbf{\xi_2}$}
\label{sec:var}
\begin{figure}[h]
\centerline{
\includegraphics[height=4.5cm,width=7cm]{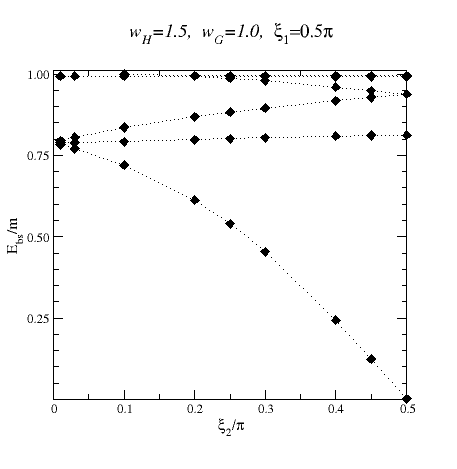}
\hspace{1.5cm}
\includegraphics[height=4.5cm,width=7cm]{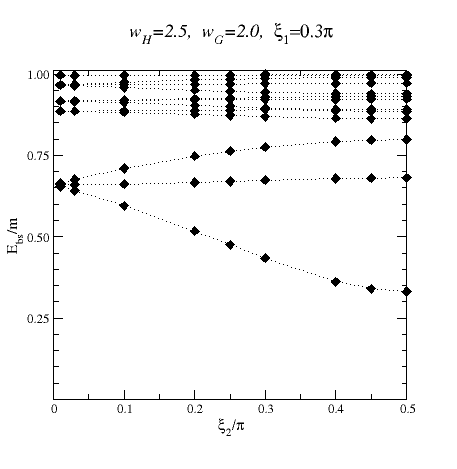}
}
\caption{\label{fig:bs}The variation of the zero modes as 
a function of $\xi_2$ for narrow (left panel) and wide (right
panel) string profiles. The dots indicate calculated
values while the lines have been added to guide the eye.}
\end{figure}

So far the numerical analysis of the charged string was limited to the case of
$\xi_2 = \pi/2$, for which the scattering and bound state problems only involve real 
valued Hamiltonians.  First numerical results for the vacuum polarization energy 
at $\xi_2 \neq \pi/2$ are shown in the right panel of figure~\ref{fig_onshell}. 
They suggest that reducing $\xi_2$ may indeed further decrease the vacuum 
polarization energy. However, the contribution from explicitly occupying the 
bound fermion levels may have the opposite effect. The numerical determination 
of the bound state energies is more complicated when the Hamiltonian is not real. 
Rather than doubling the diagonalization problem by separating real and imaginary 
parts, it is easier to extract the bound state energies from the roots of the Jost 
function at complex momentum. Since this Jost function results from a multi--channel 
problem, its roots need not be simple which adds to the difficulty of extracting 
them numerically. Fortunately, Levinson's theorem \cite{Levinson:1949,Puff:1975zz}
yields the number of expected bound states from the phase shifts. 
The typical behavior of the bound state energies as a function of $\xi_2$ is shown 
in figure \ref{fig:bs}. Note that the zero mode for $\xi_1=\frac{\pi}{2}$ 
and $\xi_2=\frac{\pi}{2}$ is recovered. Similarly to decreasing $\xi_1$, 
any reduction of $\xi_2$ also reduces the binding of that level.

The data from the $\xi_2$-variation can only lower the minimal total 
energy which further reduces the bound on $f$ and thus the mass of the 
trapped fermion below which a charged cosmic string is stable.
The conclusions made at the end of the previous section remain valid, 
but the bounds and statements become more stringent.

\section{Closed Nielsen--Olesen Strings}
\label{sec:closed}
Cosmic strings usually carry magnetic flux and are thus particle analogs of 
Abrikosov \emph{vortices} in type-II superconductors. Since the flux is 
conserved, the vortex must either end on (magnetic) charges or be closed. The 
preferred topology of strings formed during the electroweak phase transition 
are thus closed \emph{rings} which may be stabilized dynamically -- either 
by spinning at high angular momenta \cite{Witten:1984eb}, by extra currents running 
along the ring \cite{Witten:1984eb,Davis}, or by trapping charged particles as 
discussed earlier.  Such stable vortex configurations are called \emph{vortons}
\cite{Davis:1988ij}.
 
If none of these stabilizing effects is operative, the vorton will follow its 
(classical) instability and \emph{collapse} by radiating away its energy. 
Again, this picture may be altered by quantum fluctuations: Due to the 
uncertainty principle, there is a quantum mechanical penalty for localizing 
in a very small volume. This quantum energy may ultimately balance the 
classical energy of the vorton and stabilize it at some small radius.

The size of such quantum stabilized vortons is expected to be of the order 
of the Compton wavelength of the fluctuating field -- not the astrophysical 
scales over which the initial string stretched. Nonetheless, vortons are 
expected to be produced during the electroweak phase transition at such 
high rates that they would naturally evolve into acceptable densities of 
(neutral) dark matter \cite{Davis:1988ij}. 
There are also stringent experimental constraints on the density of 
charged dark matter \cite{Fukugita:1990uh} and thus charged vortons, 
which may help to rule out certain cosmology models that produce 
such vortons copiously.

In this section, we investigate a quantum mechanism to stabilize 
neutral close Nielsen--Olesen type of vortices against collapsing to 
zero radius. 
The model is that of scalar electrodynamics with the Lagrangian as in 
eq.~(\ref{eq:Lboson}), but with both the Higgs and the gauge fields now 
$U(1)$ valued. The fields of the Nielsen--Olesen string along the infinitely 
long symmetry axis are similar to those in eqs.~(\ref{eq:Higgs}) 
and~(\ref{eq:gauge}) with the $2\times2$ matrices replaced by the identity matrix. 
They are transformed to closed strings by taking the symmetry axis to
be of finite length, say $2\pi R$, and then connecting the end points.
The symmetry axis thus turns into a circle along which the core of
the string is located. When this circle is embedded in the 
$x$--$y$ plane and centered in the origin, it is appropriate to 
introduce \emph{toroidal coordinates}~\cite{book_MF} 
as\footnote{See the cited reference \cite{book_MF} for a visualisation 
of these coordinates.}
\begin{equation}
x = R\,\frac{\sinh \tau \cos\varphi}{\cosh \tau - \cos \sigma}\,, \qquad 
y = R\,\frac{\sinh \tau \sin\varphi}{\cosh \tau - \cos \sigma}\,, \qquad
z = R\,\frac{\sin \sigma}{\cosh \tau - \cos \sigma}\,.
\label{eq:tcord}
\end{equation}
The angle $\varphi\in[0,2\pi]$ parameterizes the symmetry transformation along the 
core circle, \emph{i.e.} the profiles will not depend on this variable. For a constant 
angle $\varphi$, the lines of constant $\tau \in [0,\infty]$ are circles that enclose 
the core of the string. Smaller values of $\tau$ correspond to larger circles that 
spread out towards infinity on the outside, and approach the $z$--axis on the inside. 
The inverse $1/\tau$ is therefore (in a highly non--linear fashion) related to the 
distance from the string core, similarly to the radius $\rho$ in the linear Nielsen--Olesen 
string.  Finally, the angle $\sigma\in[0,2\pi]$ parameterizes the points on the 
fixed $\tau$-circles  enclosing the string core, \emph{i.e.}~it acts like an azimuthal 
angle when viewed from the core circle. It is therefore natural to give the 
Higgs field winding  by taking $\sigma$ as its phase. Then the profiles are 
predominantly functions of $\tau$ only, independent of $\varphi$, with
(probably moderate) modulation in $\sigma$, mainly in the vicinity of the origin. 
The origin corresponds to $\tau=0$ and $\sigma=\pi$, while spatial infinity is 
approached as $\tau\to0$ and $\sigma\to0$.

Note that the $\sigma$--modulation, even if it turns out to be fairly small, cannot be 
avoided all together: For fixed $\varphi$ and $\tau=0$, $\sigma=0\ldots\pi$ parameterizes 
the positive $z$--axis. At each point along this axis the core of the string is seen 
with a different opening angle. For $\tau>0$ but $\tau \ll 1$, varying $\sigma$ corresponds 
to moving along a circle enclosing the string core. One side of this circle corresponds 
to the  far distance, while the other side is close to the origin. Obviously, different 
$\sigma$ values refer to distinct phyiscal regions and the fields must hence depend on 
this angle. In contrast to the linear Nielsen--Olesen string, the profile functions 
\emph{must} hence depend on two coordinates, which complicates matters significantly. 
To be explicit, an appropriate ansatz is
\begin{equation}
\Phi=vf(\sigma,\tau)\,{\rm e}^{in\sigma} \qquad {\rm and}\qquad
\mbox{\boldmath$A$\unboldmath}=
\frac{n}{eR}\,\eta(\sigma,\tau)\,g(\sigma,\tau)
\mbox{\boldmath$e$\unboldmath}_\sigma\,.
\label{eq:tstring}
\end{equation}
The metric factor $\eta=\cosh \tau - \cos \sigma$, as well as the inverse coupling
constant and radius, have been introduced to ensure that $f\equiv1$ and $g\equiv1$ 
correspond to a pure gauge configuration. Recall that the profiles~$f$ 
and $g$ depend on the choice of the dimensionless radius $\hat{R} \equiv m_W R$.
With these preliminaries the classical energy becomes
\begin{equation}
\frac{E}{m_W} = \frac{\pi}{g^2} 
\int_0^\infty d\tau\int_{-\pi}^\pi d\sigma\,\sinh\tau\,\Bigg\{
\frac{n^2}{\hat{R}\eta}\,(\partial_\tau g)^2 +
\hat{R} \eta\,\Big[ (\partial_\tau f)^2 + n^2 f^2 \,\big(1 - g\big)^2 +
(\partial_\sigma f)^2\Big] 
+ \frac{m_\Phi^2}{4m_W^2}\,\hat{R}^3\,\eta^3\,\big(1 - f^2\big)^2
\Bigg\}\,,
\label{eq:ECL}
\end{equation}
subject to the boundary conditions
\begin{equation}
f(0,\tau)=f(2\pi,\tau)\,,\qquad g(0,\tau) = g(2\pi,\tau)\,,
\qquad {\rm and} \qquad
 f(\sigma,\infty) = g(\sigma,\infty) = 0\,.
\label{bc1}
\end{equation}
Furthermore, Neumann boundary conditions for 
$\sigma\ne 0,2\pi$,
\begin{equation}
\partial_\tau f(\sigma,0) = \partial_\tau g(\sigma,0) = 0\,,
\label{bc2}
\end{equation}
ensures that any solution of the field equations with finite total energy 
will approach $f=g=1$ as $\sigma\to0$ or $\sigma\to2\pi$ when $\tau=0$.

\begin{figure}[t]
\centerline{
\includegraphics[height=4.0cm,width=6cm]{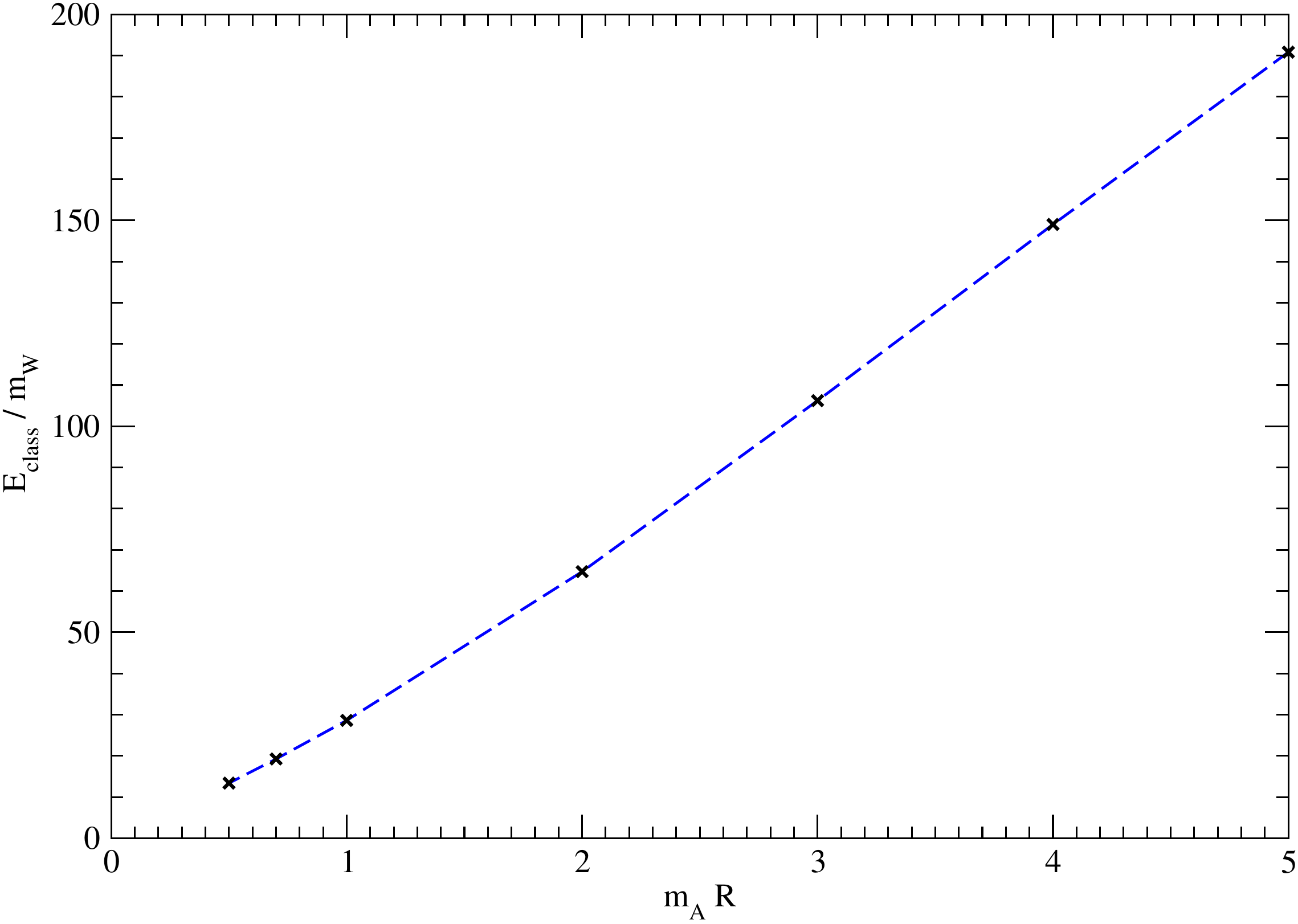}
\hspace{2cm}
\includegraphics[height=4.0cm,width=6cm]{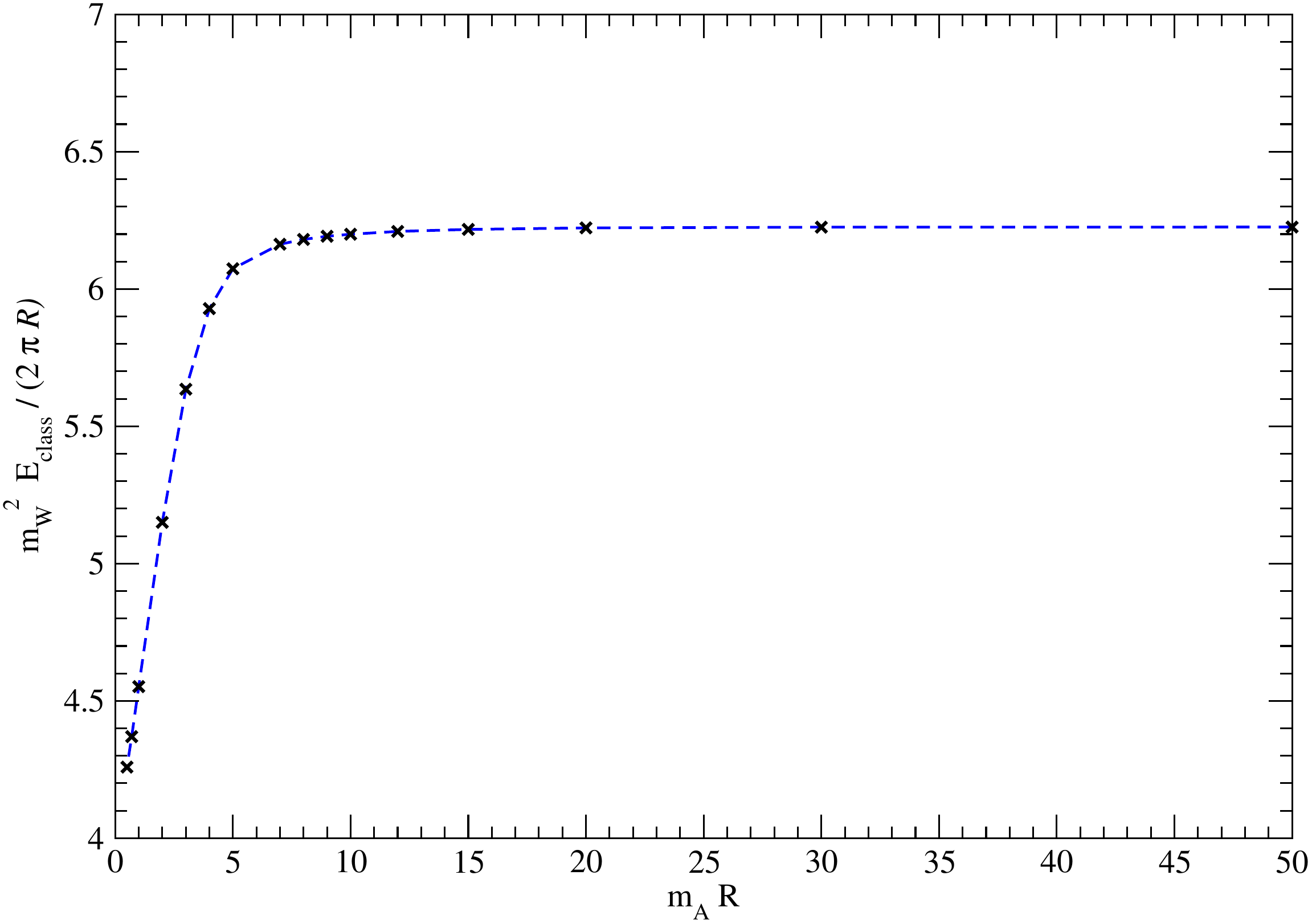}
}
\caption{\label{fig:tclass}The classical energy (left panel) 
and the energy per unit length of the core circle (right panel) 
for a closed Nielsen--Olesen string as functions of the core
radius. The stars indicate the numerically computed cases. Data 
are presented for $g= 1 / \sqrt{2}$ and $m_\Phi=m_W$, for which the 
straight string has energy per unit length $6.28\,m_W^2$.}
\end{figure}

A detailed description of the relaxation method for constructing the actual 
field configuration numerically constrained to a prescribed value of the 
radius parameter $\hat{R}$ can be found in ref.~\cite{Quandt:2013qxa}.
Here we just review the results for the minimal energy in 
figure~\ref{fig:tclass}.
The left panel essentially reveals an almost linear relation between 
the minimal energy and the radius. More interesting is the energy 
per length of the torus, which is shown in the right panel. 
The first observation is a saturation at large radii. This is nothing but the 
energy per unit length of the infinitely long string, \emph{i.e.} the 
Nielson-Olesen limit. Reproducing this known result is a strong consistency 
check of our numerical treatment. Interestingly, the saturation 
occurs already for radii only slightly larger than the Compton wave--length of 
the gauge field. This suggests that energy loss from radiating gauge and/or Higgs 
fluctuations is indeed irrelevant for simulations of networks of closed strings 
that refer to cosmological length scales. For smaller radii, the interaction 
between opposite sides of the torus leads to a considerable drop in the energy 
per unit length. As a consequence, the total energy of the torus configuration 
decays more than linearly at small core radii $R\to0$, which means that it is 
classically unstable against shrinking to a point. Of course, this instability 
is expected from Derrick's theorem, according to which higher order derivative 
terms (as in the Skyrme model\cite{Skyrme:1961vq}) would be needed 
to stabilize a localized object in three space dimensions.

\begin{figure}[t]
\centerline{
\includegraphics[height=4.0cm,width=6.5cm]{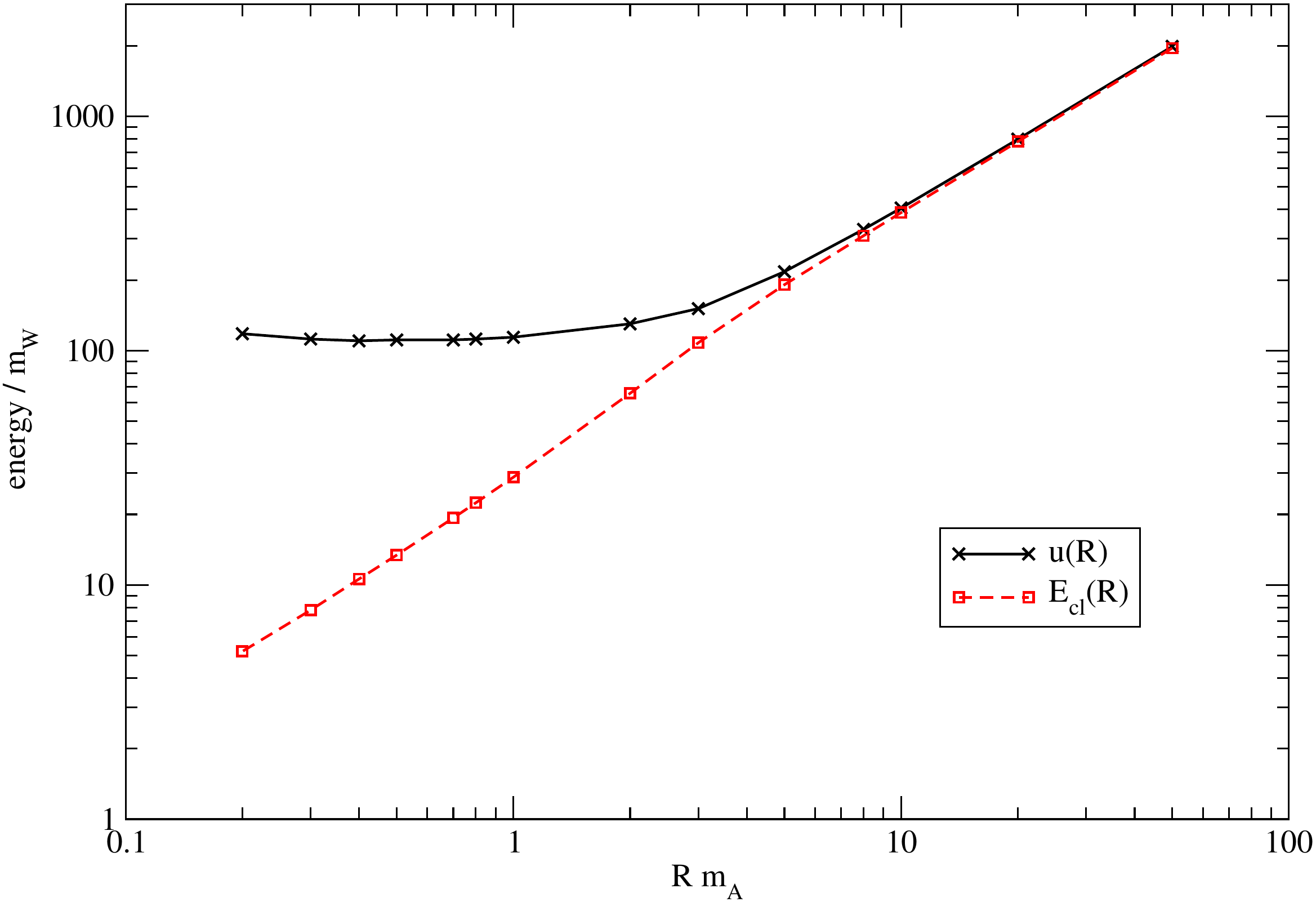}
\hspace{1.5cm}
\includegraphics[height=4.0cm,width=6.5cm]{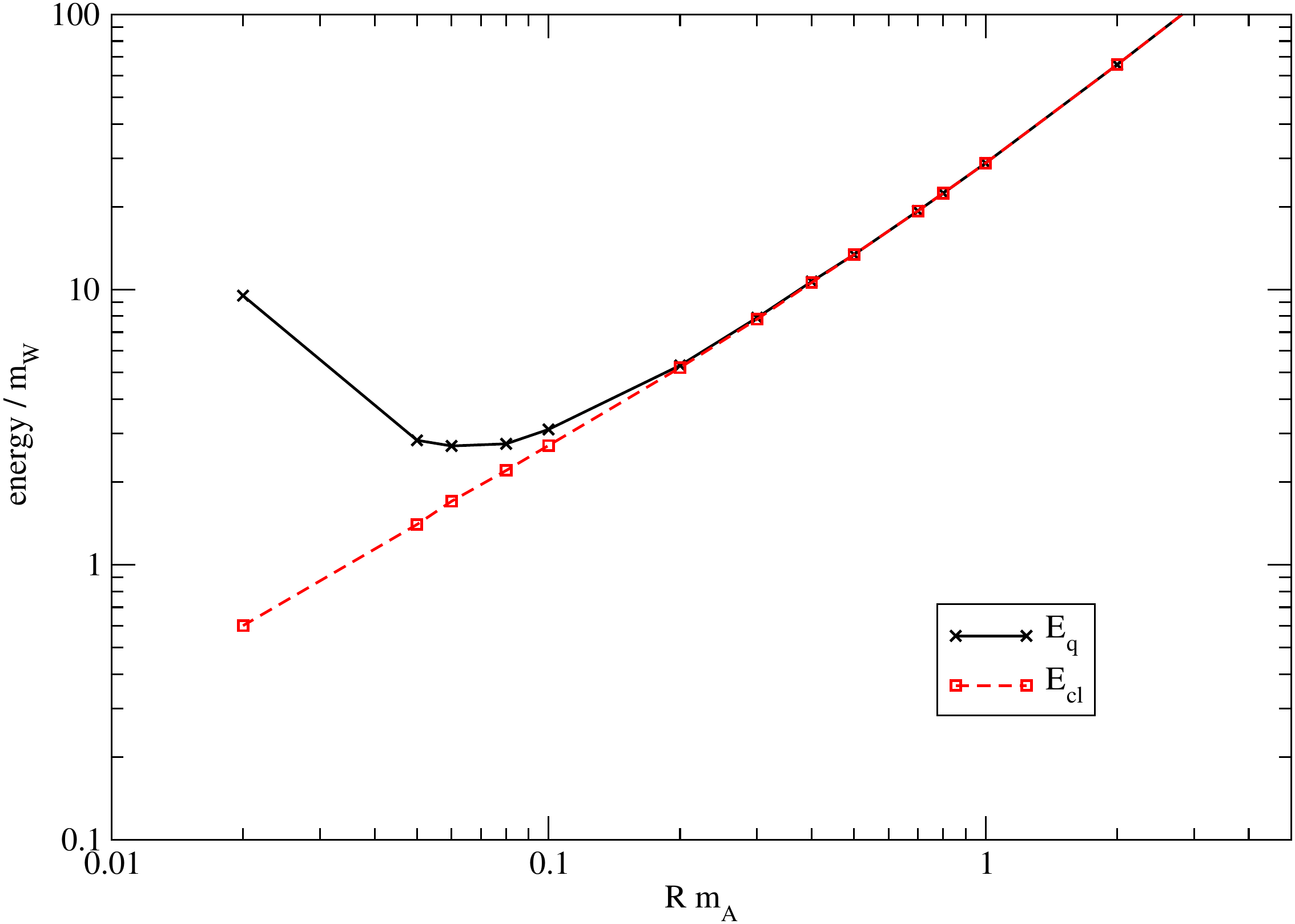}
}
\caption{\label{fig:Eqtor}The mass functional for the unstable mode (left panel) and 
the lower boundary for its quantum energy (right panel). Also shown are the classical 
energies.}
\end{figure}

Of course, quantum mechanically such an instability could represent a 
violation of Heisenberg's uncertainty principle according to which 
field configurations localized within an arbitrarily small region 
must have large energy and/or momentum. To investigate this question, we 
elevate the radius to quantum variable and treat it as a collective coordinate. 
In a first step we take $R=R(t)$ to be time dependent and then identify its 
conjugate momentum. Since the Cartesian coordinates are fixed in terms of $R$, 
this implements a time dependence of the fields via eq.~(\ref{eq:tcord}). Once 
$\mbox{\boldmath$A$\unboldmath}$ is time dependent, its temporal component
can no longer vanish but will be induced via Gau{\ss}' law. As a result,
it will be proportional to the time derivative $\dot{R}=\frac{dR(t)}{dt}$. 
Thus the Gau{\ss}' law augments the field configuration by the 
additional profile function $q$ via
\begin{equation}
e\,A_0(\mathbf{x},t)= \frac{\dot{R}}{R}\,q(\sigma,\tau)\,.
\label{eq:a0}
\end{equation}
Putting things together yields the action functional for the unstable mode
\begin{equation}
S = \int dt\,\left[ \frac{1}{2}\,u(R)\,\dot{R}^2 - E(R) \right]\,.
\label{eq:qaction}
\end{equation}
The energy functional $E(R)$ is that of eq.~(\ref{eq:ECL}) while the explicit 
expression for the {\it mass} functional $u(R)$ is fairly complicated and given in
detail in ref.~\cite{Quandt:2013qxa}. Here it suffices to mention that $u(R)$ 
contains linear and quadratic terms in the induced profile $q$. The 
profile functions $f$ and $g$ are determined from minimizing $E(R)$.
In turn they provide source terms
for $q$ when applying the variational principle to $u(R)$. This procedure
is equivalent to solving the Gau{\ss} law constraint~\cite{Quandt:2013qxa}.
The resulting mass functional is shown in the left panel of 
figure \ref{fig:Eqtor} as a function of the radius $R$. 

The action functional, eq.~(\ref{eq:qaction}) yields the canonical
momentum (operator) $p$ and, via Legernde transformation, the (quantum) energy of the 
unstable mode as
\begin{equation}
p=u(R)\frac{dR}{dt}\qquad {\rm and} \qquad
E=\frac{p^2}{2u(R)}+E(R)\,.
\label{eq:Eqtor}
\end{equation}
Figure \ref{fig:Eqtor} shows that $u(R)$ is huge for radii larger than the Compton 
wave--length of the gauge boson. For a fixed momentum, the back--reaction of the 
classical profiles $f$ and $g$ due to the induced kinetic term in eq.~(\ref{eq:Eqtor})
is of the order $\mathcal{O}(u^{-2})$. 
This effect can thus be safely omitted and these profile functions
are reliably obtained by applying the variational principle to $E(R)$ alone. 
As mentioned earlier, $q$ is subsequently obtained from varying $u(R)$, 
and, since initial profiles $f$ and $g$ need not be re-computed, 
the parameter functions in eq.~(\ref{eq:Eqtor}) are then completely determined.
The corresponding Hamilton operator can then be constructed by imposing canonical 
commutation relations\footnote{The metric factors in the curvilinear coordinates
lead to ordering ambiguities which require particular care.}
for $p$ and $R$. The ground state (g.s.) energy eigenvalue can easily be estimated 
by recalling that the uncertainty principle establishes its lower bound in the form
\begin{equation}
E_{\rm g.s.}\ge\frac{\hbar^2}{2u(R)R^2}+E(R)\,.
\label{eq:Heisenberg}
\end{equation}
This total energy of the torus string is displayed
as a function of $R$ in the right panel of figure \ref{fig:Eqtor}. Obviously a 
local minimum at non--zero torus radius $R$ exists, suggesting the quantum 
stabilization of the closed Nielsen--Olesen string. As for the infinitely 
long charged string, the length scale at which stability occurs is quite small 
and it is unclear whether it leads to significant cosmological implications.

\section{Conclusion}
\label{sec:conclusion}
This review has mainly focused on computing the fermion
contribution to the vacuum polarization energy per unit length of an 
infinitely long straight string in a simplified version of the electroweak 
standard model. The approach is based on the interface formalism, for 
which the analytical properties of scattering data are essential. Obstacles
that arise in a na\"{\i}ve treatment from the non--trivial structure of
the string configuration at spatial infinity are circumvented by choosing
a particular subset of gauges. Numerically 
the vacuum polarization turns out to be small and positive 
in the regime in which the one--fermion loop approximation is reliable. 
Hence, there is no quantum stabilization of the (uncharged) string. 
However, it also turns out
that a heavy fermion doublet \emph{can} stabilize a non--trivial string background 
if such fermions are trapped at the string core to give it a non--zero fixed 
charge per unit length. The resulting configuration is dominated by the Higgs 
field. Since any additional variational degree of freedom can only lower the 
total energy, the embedding of this configuration  in the full standard model, 
with the $U(1)$ gauge field included, will also yield a bound object, at least 
if the mixing between this heavy fermion and the known standard model 
fermions can be ignored. Numerically binding sets in at 
$m \approx 300\,\mathrm{GeV}$, which is still within the range of energy scales 
at which the standard model is expected to provide an effective description of the 
relevant physics, and also within the range to be probed at the LHC. Light fermions 
would contribute only weakly to the binding of the string, since their Yukawa couplings 
are small.  We emphasize that this cosmic string configuration is a novel solution 
in a model that is very closely related to the standard model of particle 
physics. At a similar scale, though through a different mechanism, quantum effects 
may also prevent neutral closed strings from collapsing.

\section*{Acknowledgments}

This {\it brief review} is based on a presentation by H.W. at the 
{\it 4th winter workshop on non--perturbative quantum field theory}, 
Antipolis (France), Feb. 2015. The work of H.W. is founded in parts by the 
National Research Foundation NRF,  grant~77454. N.G. was supported in 
part by the National Science Foundation (NSF) through grant PHY-1213456.

\pagebreak



\end{document}